# Room-Temperature Terahertz Photoconductivity Polarity Switching in High Entropy Nickelates with Implications for Photonic Synapses

Sanjeev Kumar, Brijesh Singh Mehra, Gaurav Dubey, Prakhar Vashishtha, Neeraj Bhatt, Jayaprakash Sahoo, Ravi Shankar Singh and Dhanvir Singh Rana[*]

S. Kumar, B.S. Mehra, G. Dubey, P. Vashishtha, N. Bhatt, J. Sahoo, R.S. Singh, D.S. Rana
Department of Physics
Indian Institute of Science Education and Research Bhopal, Bhopal 462066, Madhya Pradesh, India
E-mail: dsrana@iiserb.ac.in

## Abstract

High entropy oxides (HEO) hold the potential to revolutionize the conventional material paradigms by leveraging high order of chemical disorder that induces highly desirable exotic phases for advanced applications. Here, we devise a methodology to enhance the efficiency of an artificial photonic synapse using a high entropy rare earth nickelate. Combined with epitaxial strain, we show that high entropy can further manipulate the phase of these locally disordered materials. Using time-averaged and time resolved Terahertz (THz) spectroscopy as dynamic probe, for the first time we show a rare combination of i) crystal axis dependent insulator to metal THz electronic phase transition and ii) coexistence of negative and positive THz photoconductivity at room temperature. Detailed analysis within theoretical models, including density functional theory (DFT)-based band structure calculations, suggest origin of these properties as disproportionate ordering of oxygen vacancies. Based on these findings, a conceptual THz-based artificial photonic synapse is proposed. This work underlines the pivotal role of HEO in advancing diverse THz functionalities, representing a critical step toward futuristic applications like THz-based high- speed computing and communication with an emphasis in THz frequency domain.

# Introduction

The growing demand for high-speed computing in artificial intelligence has spurred interest in brain-inspired neuromorphic architectures that emulate biological neural systems.[1] Among the key components—artificial neurons, synapses, spiking networks, and memristors—artificial synapses play a crucial role by modulating synaptic strength to enable learning and signal transmission.[2] While electronic synapses have been widely explored,[3–6] over 70% of human sensory input originates from vision,[7] prompting a surge of interest in photonic synapses that mimic the visual perception.[8] Also, integrating photonic control enhances energy efficiency, speed, and reliability in neuromorphic systems.[9,10] Achieving reversible optical writing and erasing within a single photonic synapse is essential for realizing fully optical, energy-efficient neuromorphic operation. Although recent advances—such as self-powered $CsPbBr_3$/carbon nitride,[9] spiropyran-based organic,[11] and hybrid heterostructure systems[12,13] —demonstrate rapid light-driven memory modulation, they still depend on separate wavelengths or hybrid optical–electrical stimuli for writing and erasing.

Realizing an efficient, reversible artificial photonic synapse requires carefully engineered materials and precise methodologies to enable controllable optical responses. The quest for suitable materials is driven by the ability to tune lattice, spin, charge, and orbital degrees of freedom for novel functionalities.[14] Chemical substitution, for instance, can restructure lattice sites and give rise to fascinating phases such as superconductivity,[15] metal-insulator transitions, and colossal magnetoresistance,[16] underscoring its significance in both fundamental and applied research. Based on this lattice reconfiguration strategy, the high entropy approach offers a powerful route to explore heavily cation-doped materials and uncover emergent functionalities. High entropy materials consist of five or more elements at a cation site in equimolar concentration,[17] resulting in a wide variety of possible atomic arrangements, thus, yielding a high configurational entropy of mixing ($\Delta S_{mix} = - K_B \ln\Omega$, where $K_B$ is Boltzmann constant and $\Omega$ is number of possible microstates).[18] This mixing entropy contributes to the Gibbs free energy ($\Delta G_{mix} = \Delta H_{mix} - T \Delta S_{mix}$, where $\Delta H_{mix}$ is enthalpy of mixing, T is temperature and $\Delta S_{mix}$ corresponds to entropy of mixing) and overcomes the enthalpy driven phase separation, facilitating the formation of single structural phase.[17] Initially developed for alloys,[19,20] high entropy concept has been extended to oxides [21] unravelling exceptional properties such as, high ionic conductivity,[22] emergent magnetic ordering, [23,24] highly anisotropic thermal expansion,[25] etc. Yet, their microscopic complexity and functional potential remain underexplored. The inherent compositional disorder and tunable electronic structure of high entropy oxides (HEO) provide an ideal platform for manipulating carrier dynamics, dielectric behavior, and light–matter interactions—key factors for advanced device applications.

We investigated HEO in the terahertz (THz) region of the electromagnetic spectrum, this enables ultrafast, coherent, broadband, and energy-efficient manipulation of their complex optoelectronic states, which can be useful for realization of all-photonic neuromorphic device components. THz spectroscopy, an emerging and highly sensitive probe, enables direct access to the intertwined electronic, magnetic, and structural energetics arising from local compositional and lattice variations. Its low photon energy allows detection of nearly degenerate energy scales without perturbing the intrinsic state of the system, establishing its relevance across diverse domains such as computation,[26,27] communication,[28] medicine,[29] etc. Rare-earth nickelates, known for their multiband correlated behavior, develop complex energy landscapes when highly disordered at the A-site in their HEO form.[30] Their intermediate ("bad-metal") conductivity and strong THz sensitivity to carrier dynamics make them ideal for tunable THz response studies, offering pathways for controllable THz modulation and next-generation ultrafast optoelectronic devices.[31]

In this study, we report the THz investigations on high entropy nickelate thin films of $(La_{0.2}\ Pr_{0.2}\ Nd_{0.2}\ Sm_{0.2}\ Eu_{0.2})\ NiO_3$, under varying strains and different crystal symmetries. Time domain spectroscopy uncovered an anisotropic THz conductivity, primarily influenced by oxygen vacancies. Optical pump–THz probe (OPTP) measurements revealed a rare phenomenon—reversible switching from negative to positive THz transmittance on (001) oriented NdGaO₃ substrates upon azimuthal rotation at room temperature. This, along with a transient reversal of THz photoconductivity on a few picoseconds' timescale form a unique set of properties for designing a THz-driven artificial photonic synapse, offering a promising platform for future high-speed neuromorphic computing technologies.

## Results and Discussion

Thin films of high entropy $(La_{0.2}\ Pr_{0.2}\ Nd_{0.2}\ Sm_{0.2}\ Eu_{0.2})\ NiO_3$ [(LPNSE)NO] were synthesized on (100) oriented cubic $LaAlO_3$ [(LPNSE)NO/LAO(100)] and (001) oriented orthorhombic $NdGaO_3$ [(LPNSE)NO/NGO(001)] substrates which are coherently compressive and tensile strained respectively [structural data appended in Supplementary Information, S1]. Electronic transport properties of these epitaxial films reveal a metallic nature at room temperature [Supplementary Information, S2] with the resistivity of (LPNSE)NO/NGO(001) greater than (LPNSE)NO/LAO(100). Both high entropy films exhibited a metal to insulator transition along with hysteresis during cooling and heating of sample - a characteristic feature of rare earth nickelate thin films. Direct current (DC) electronic transport measurement, limited by contact resistance, provides a measure of the overall conductivity. In contrast, THz spectroscopy, through its sensitivity to intrinsic electric and magnetic dipole responses, offers access to the microscopic charge carrier dynamics [ representative raw THz data available in Supplementary Information, S3].

**Microscopic insights through THz based studies**

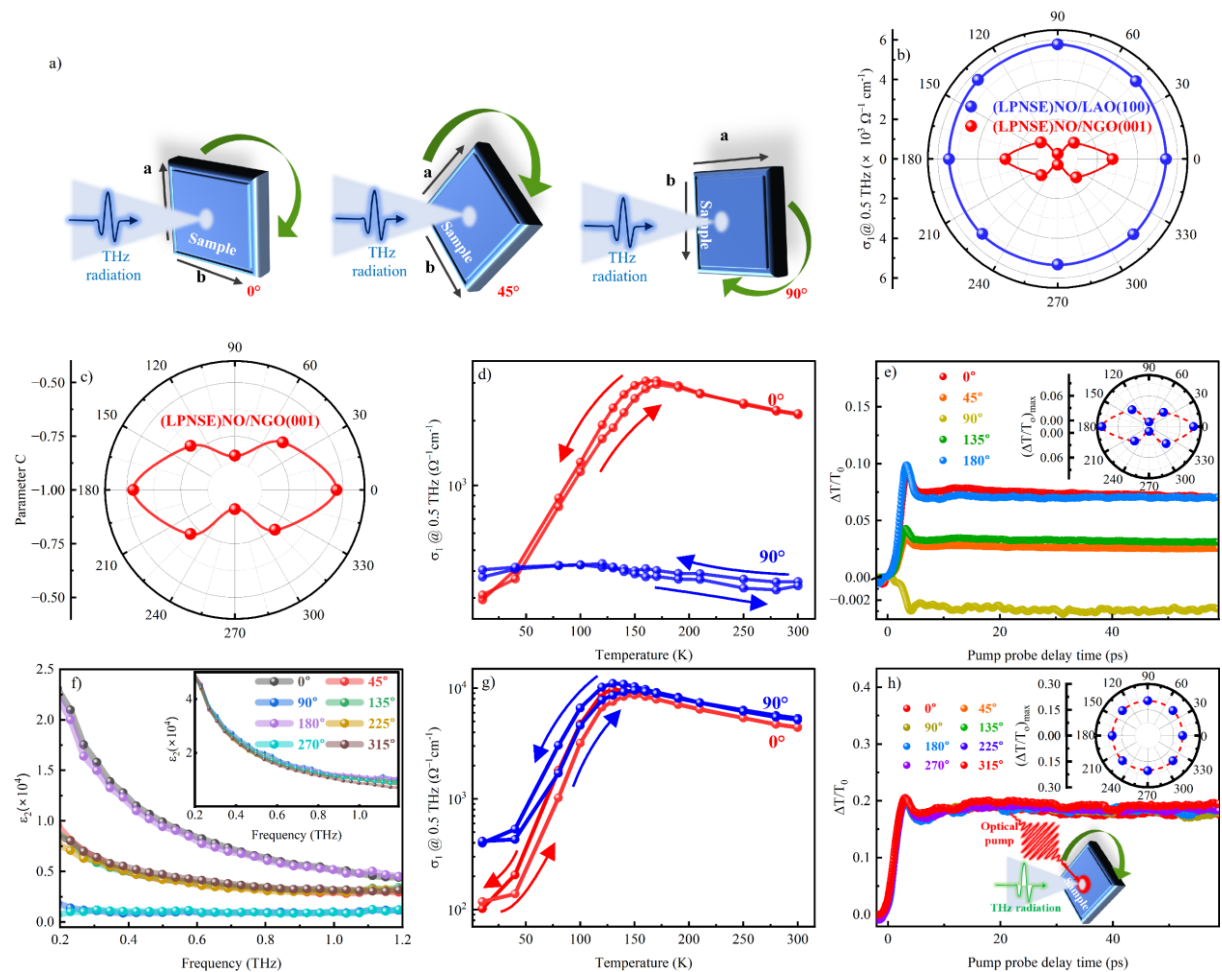

Figure 1: Exploring azimuth dependence in time-averaged and time-resolved THz spectroscopy. (a) Scheme of sample rotation for azimuth dependent measurements. (b) Real THz conductivity for (LPNSE)NO/LAO(100) and (LPNSE)NO/NGO(001) thin films at 0.5 THz frequency and different sample azimuths. (c) Variation of C parameter with sample azimuth for (LPNSE)NO/NGO(001). (f) Imaginary dielectric function at different azimuths for (LPNSE)NO/NGO(001), inset shows same for (LPNSE)/LAO(100). Temperature dependent real THz conductivity at 0.5 THz along orthogonal crystal axes for (d) (LPNSE)NO/LAO(100) and (g) (LPNSE)NO/NGO(001) thin films. (e) and (h) illustrates transient THz transmission at various sample azimuths

Incorporation of multiple elements at A-site and the crystallographic mismatch between the substrate and the high entropy nickelate layer can give rise to anisotropic epitaxial strain, which in turn may induce pronounced direction-dependent physical properties along distinct crystallographic axes. Hence, time-domain THz spectroscopy was carried out along different in-plane crystallographic orientations by rotating the sample about its azimuthal axis, as illustrated in Figure 1(a). The results reveal lower THz conductivity for (LPNSE)NO/NGO(001) compared to (LPNSE)NO/LAO(100), in accordance with the DC conductivity [Supplementary Information, Figure S2]. A pronounced anisotropy in THz conductivity was observed across different in-plane orientations (sample azimuths) in the tensile-strained film, this, in contrast, is absent in the compressively strained film, as shown in Figure 1(b) [detailed representative data in Supplementary Information, Figure S4]. The variation in conductivity with sample azimuth can be a result of preferential ordering of regions/domains with different conductivities. Therefore, the dielectric function was analysed under the Maxwell-Wagner (MW) relaxation framework, which accounts for the system's heterogeneity [details in Supplementary Information, S5]. MW behaviour, commonly observed in heterogeneous systems such as composites, metal/insulator interfaces, metal/semiconductor boundaries, etc., is characterized by a significant increase in imaginary dielectric function ($\varepsilon_2$) as the frequency approaches zero. In (LPNSE)NO/LAO(100), $\varepsilon_2$ followed the same trend across sample azimuths [inset, Figure 1(f)], indicating uniformity. In contrast, (LPNSE)NO/NGO(001) displayed significant variation with sample azimuth [Figure 1(f)], confirming anisotropic heterogeneity.

To elucidate the underlying mechanism of the observed anisotropic behaviour, a comprehensive analysis of THz conductivity was performed under the framework of Drude-Smith model [32] by fitting THz conductivity on equation (1) [representative fitting presented in Supplementary Information, S6].

$$\sigma = \frac{\varepsilon_0 \omega_p^2}{\tau - i\omega} \left[ 1 + \sum_n \frac{c_n \tau}{\tau - i\omega} \right] - i\varepsilon_0 \omega \, (\varepsilon_\infty - 1) \qquad (1)$$

Here, '$\omega_p$' is plasma frequency, '$c_n$' represents fraction of electron's velocity retained after nth collision, '$\varepsilon_0$' and '$\varepsilon_\infty$' are permittivity of free space and permittivity at higher frequency respectively, '$\tau$' is the scattering rate and '$\omega$' is the frequency. To simplify, we have considered the "persistence of velocity" parameter for the case of one collision only ($C_n= C$), parameter 'C' can acquire values from 0 to -1, where the larger negative value indicates more disorder and vice-versa.[33] Figure 1(c) presents the 'C' values obtained by fitting the THz conductivity spectra on equation (1) for (LPNSE)NO/NGO(001) at various sample azimuths, which follows the behaviour of steady-state conductivity across all angles [Figure 1(b)]. A more negative 'C' value is associated with greater backscattering which may be caused by oxygen vacancies, grain boundaries and/or other crystal defects.[34,35] However, given that the thin films used in this study are of high quality and exhibit epitaxial coherence, the contributions from the latter two possibilities can be ruled out. Therefore, observed variation in the 'C' parameter with sample azimuth is attributed to an asymmetric distribution of oxygen vacancies along the crystallographic axes. In-plane epitaxial tensile strain also promotes the formation of oxygen vacancies in rare earth perovskite thin films [36] by reducing electron-electron repulsion along the broken B-O-B bond axis, thereby decreasing the vacancy formation energy.[37] In this context, non-uniform tensile strain along different in-plane crystallographic axes can induce an inhomogeneous distribution of oxygen vacancies, resulting in anisotropic conductivity.

The interplay between high doping levels and epitaxial strain in nickelate system can potentially drive the emergence of novel or unconventional phase transitions. To elucidate this, temperature dependent THz conductivity was examined at different azimuths for both tensile and compressively strained films [Figure 1(d,g), peak amplitude vs temperature are appended in Supplementary Information, S7 and representative THz conductivity data in Supplementary Information, S8]. The compressively strained film exhibits identical behaviour along orthogonal in-plane axes, showing a well-defined insulator-to-metal transition with hysteresis [Figure 1(g)]. In contrast, the tensile-strained film exhibits an electronic transition with hysteresis along the 0° azimuthal angle, while the 90° azimuthal angle displays anomalously low THz conductivity, with no indication of a transition [Figure 1(d)]. This observation has two key implications. Fundamentally, it provides strong evidence for anisotropic ordering of oxygen vacancies in the film. Specifically, at room temperature, the conductivity along the 90° axis is lower [Supplementary Information, Figure S14(ii)], consistent with enhanced carrier scattering from oxygen vacancies acting as localized defect centers. However, upon cooling, the conductivity along this axis becomes relatively higher—a trend that can be understood as oxygen vacancies acting as donor and decreasing the film's resistivity.[38] The observed crossover—from a clear metal-insulator transition behavior to its suppression upon 90° rotation—indicates that oxygen vacancies possess a preferential alignment rather than a random distribution, leading to the emergence of directional transport anisotropy. From a technological perspective, coexistence of metallic and insulating responses along orthogonal directions offers a unique platform for realizing anisotropic multistate switching devices, wherein different resistance states can be accessed by tuning the measurement direction or external perturbations. Overall, THz spectroscopy of the high entropy nickelate, studied under both compressive and tensile strain as a function of sample azimuth and temperature, suggests following scenario. Anisotropic in-plane tensile strain causes oxygen vacancies to preferentially align along the direction experiencing greater tensile strain, leading to higher charge carrier scattering. This anisotropy in scattering is reflected in the azimuth-dependent variation of backscattering parameter obtained from Drude-Smith analysis of the conductivity spectra [Figure 1(c)]. As a result, the system exhibits pronounced in-plane anisotropic THz conductivity.

**Non-equilibrium carrier dynamics**

We further investigate the influence of distribution of vacancies in excited state using OPTP spectroscopy [Supplementary Information, S9]. Normalized transient THz transmission change at the peak of transmitted THz pulse ($\Delta T/T_0$, where $\Delta T$ is the transmission change and $T_0$ is transmission in unpumped state) [details in Supplementary Information, S10], which is qualitatively a negative counterpart of real THz photoconductivity,[39,40] was extracted as a function of pump-probe delay time [representative spectrum of THz photoconductivity in Supplementary Information, S11]. A positive THz transmittance ($\Delta T/T_0$) for both the tensile and compressive films [Supplementary Information, S12] may be attributed to enhanced carrier scattering as pump beam raises the temperature of system. Transmittance further increases with increasing pump beam fluence, suggesting that the additional energy delivered by the optical beam enhances the scattering of charge carriers resulting in reduced conductivity, and hence increases THz transmission.

The azimuth dependent $\Delta T/T_0$ of tensile strained film exhibited anisotropy along with a persistent reversal of sign at a rotation of 90° [Figure 1(e)]. This is the first observation of coexisting negative and positive THz photoconductivity at room temperature in a HEO, achieved solely through azimuthal rotation. As this system is intrinsically isotropic, the most plausible explanation for the observed in-plane anisotropy in $\Delta T/T_0$ is likely the non-uniform ordering of oxygen vacancies. Observation of persistent negative $\Delta T/T_0$ response suggests the formation of free charge carriers induced by inter-band optical excitation,[41] therefore a possibility of anisotropic band structure cannot be ruled out. This feature, in contrast, is absent for compressive film [Figure 1(h)]. Additionally, few bumps were evident in fluence dependent $\Delta T/T_0$ at high fluences at particular time intervals (between 10-20 ps) [Figure S12], which can be ascribed to acoustic phonon modulations of the THz probe beam.[42,43] Decay of the THz transmittance at the timescale >100 ps points towards phonon-phonon relaxation of the heat energy acquired due to optical excitation.[44]

Oxygen vacancies in rare earth nickelates modify the structure and act as electron donors. These electrons localize on the $NiO_6$ octahedra inducing an insulating Mott state.[45] Depending on the density and distribution of vacancies, Ni local symmetry can be lowered from octahedral to square pyramidal, square planar and tetrahedral structures which drastically alters crystal field splitting. Kotiuga et al. studied this scenario and conjectured that oxygen vacancies lead to conversion of $Ni^{3+}$ to $Ni^{2+}$ reducing the hybridization of unoccupied Ni 3d states with O 2p states. This conversion also involves shifting of energy states from bottom of conduction band to top of valence band. Remaining unoccupied states associated with $Ni^{2+}$ shift up in energy, creating gap between spin-up and spin-down $e_g$ states due to onsite electron correlations and result in reduced metallic conductivity. This scenario suggests restructuring of bands at oxygen vacancy positions, and anisotropic oxygen vacancies' arrangement would imply anisotropic band structure. To buttress this proposition for the present case, density functional theory (DFT) calculations were carried out on the HEO nickelate. DFT calculations reveal a decrease in density of states at the Fermi level with increasing oxygen vacancy concentration [Figure 2(a-c)], indicative of reduced electronic conductivity [see Methods section for details on computational methodology]. Reduction in conductivity following the deliberate introduction of oxygen vacancies in the HEO structure supports our hypothesis that the anisotropic conductivity originates from preferential directional ordering of these vacancies [Figure 2(d)]. To further substantiate our claim, we synthesized a new (LPNSE)NO/NGO(001) thin film with a reduced concentration of oxygen vacancies, as confirmed by resistivity measurements [Supplementary Information, S13 (a)]. Notably, this film exhibits no anisotropy in either steady-state THz conductivity or THz transmittance upon optical excitation [Supplementary Information, S13 (b,c)]. A detailed discussion on anisotropic ordering of oxygen vacancies with quantitative evidences is appended in Supplementary Information, S14.

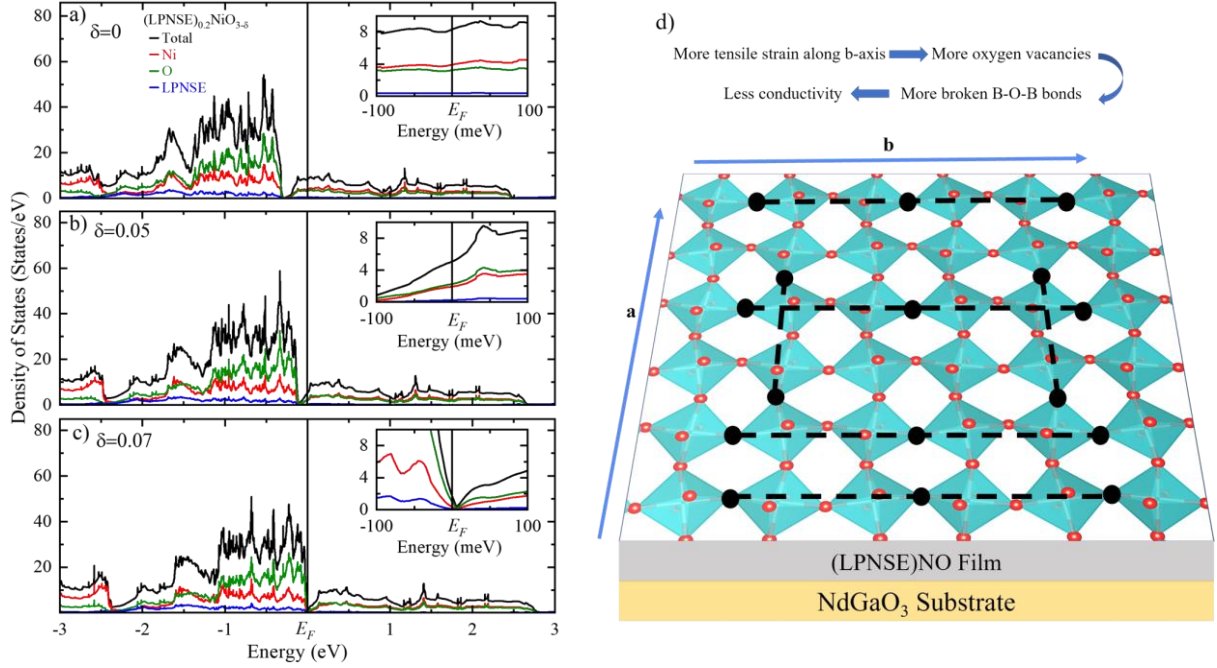

*Figure 2: Density of states for (LPNSE)NO having oxygen vacancy of (a) 0%, (b) 1.67% and (c) 2.33% (insets show intensity in the vicinity of Fermi level). (d) Representation of inequivalent oxygen vacancies' concentration (black circles) along orthogonal crystal axes leading to anisotropic conductivity.*

**Investigation of unconventional symmetry: (111) orientation**

While the role of epitaxial strain has been investigated, the influence of the underlying substrate's crystal symmetry allows another control of locally disordered high entropy nickelates. The (111) orientation of NdGaO$_3$ [NGO (111)] is known to impose tensile strain along with a higher epitaxial constraint on nickelate film as compared to conventional (001) and (100) orientations, owing to the connection of BO$_6$ octahedra of the film and substrate through 3 oxygen atoms.[46] As a result, NGO (111) substrate can prove to be a good platform to tailor the functionalities of inherently locally distorted HEO. Accordingly, (LPNSE)NO films were grown on NGO (111) [structural characterizations in Supplementary Information, S1] for THz studies. Room temperature THz conductivity obtained for (LPNSE)NO/NGO(111) exhibits lesser values than (LPNSE)NO/NGO(001) film for 0° sample azimuth [Supplementary Information, S15]. Azimuth dependent THz conductivity exhibits anisotropy similar to that of (LPNSE)NO/NGO(001) but with a different symmetry [Figure 3(a)]. Imaginary dielectric constant ($\varepsilon_2$) exhibits MW type of behaviour for all angles with varying magnitude. Contrary to the case of (LPNSE)NO/NGO(001), $\varepsilon_2$ for (111) sample does not exhibit a crossover from MW to non-MW behaviour [Figure 3(b)], which confirms that the variation in heterogeneity is lesser as compared to that in (LPNSE)NO/NGO(001).

ΔT/T₀ obtained from OPTP exhibited a peculiar behaviour in the form of transient reversal of sign from positive to negative in few picoseconds timescale [inset Figure 3(c)] and strength of both positive and negative ΔT/T₀ increased with optical pump fluence. Azimuth-dependent ΔT/T₀ demonstrated anisotropic behaviour [Figure 3(d)]. However, this anisotropy is also not as symmetrical as it was in case of (LPNSE)NO/NGO(001), which can be ascribed to the asymmetric in-plane orientation/ structure/ bonds of (111) oriented substrate. Also, no persistent reversal of ΔT/T₀ sign was observed at any angle for (LPNSE)NO/NGO(111) which is consistent with absence of non-MW type of behaviour.

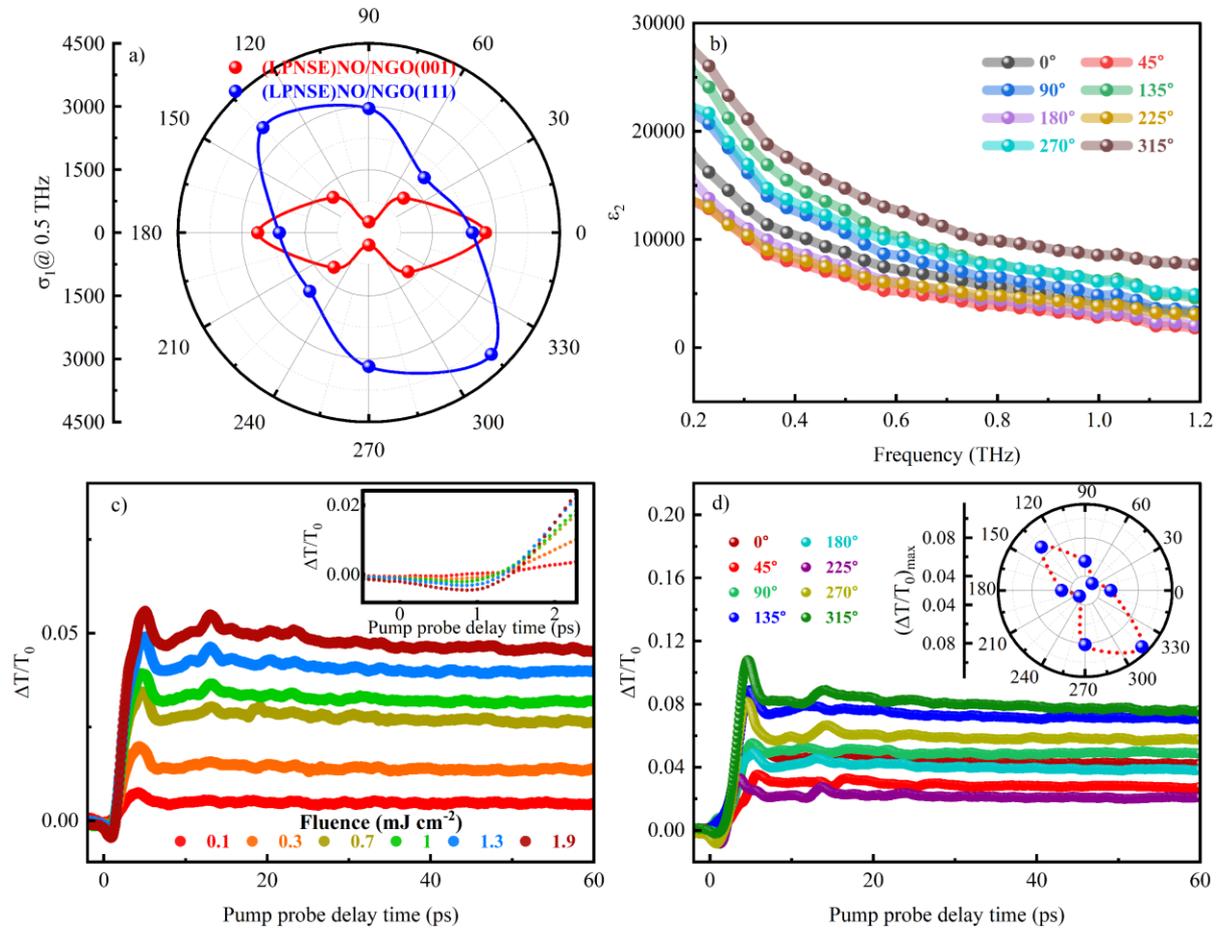

Figure 3: Time-averaged and time-resolved THz studies on unconventional (111) oriented high entropy nickelate. (a) THz conductivity as a function of sample azimuth at 0.5 THz for (LPNSE)NO/NGO(001) and (LPNSE)NO/NGO(111). (b)Imaginary dielectric function at different azimuths for (LPNSE)NO/NGO(111). (c) THz transmission change as a function pump-probe delay time at different fluences for (LPNSE)NO/NGO(111), inset shows the negative transmission change in enlarged form. (d) THz transmission change for (LPNSE)NO/NGO(111) as a function of pump probe delay time at different azimuths, inset shows maximum THz transmission as a function of sample azimuth.

The investigation of high entropy nickelates using THz spectroscopy proved to be a powerful approach to gain microscopic insights into their inherently locally distorted structures. The interplay of local disorder with tuneable parameters such as epitaxial strain, crystallographic symmetry, and sample azimuth offers a pathway to engineer enhanced optical functionalities that surpass those of their parent nickelates [detailed discussion in supplementary information, S16]. Notably, the persistent and reversible modulation of THz transmittance sign via optical control highlights the potential of these materials for integration into photonics-based devices and computing components, with implications for enhancing their overall performance and energy efficiency.

## A proposed artificial photonic synapse

Brain inspired neuromorphic computing is poised to make a significant impact in the computing landscape, driven by the growing demands of artificial intelligence-based devices and functions, as they can provide higher computing speeds compared to computers based on von-Neumann architecture.[47] These systems consist of key components such as artificial neurons, artificial synapses, spiking neural networks, memristors, and sensors. Among these, artificial synapses play a crucial role, mirroring the function of biological synapses by connecting neurons, facilitating signal transmission, and enabling learning through changes in synaptic strength.[2,48] Artificial synapses can be stimulated through various methods, including optical, electronic, magnetic, voice, chemical, temperature, and pressure, with corresponding electronic and optical responses.[49] Optically driven artificial synapses offer distinct advantages, such as high energy efficiency, rapid response times, and enhanced reliability. [9,10]While some optically driven synapses can perform writing operations using optical signals, there is a notable lack of synapses capable of erasing through optical stimulus.[50] Developing an artificial synapse that can execute both writing and erasing using optical signals would significantly enhance processing speed and energy efficiency. In this context, leveraging the coexisting negative and positive $\Delta T/T_0$, as in (LPNSE)NO/NGO(001) thin film, integrated with a non-volatile memory device,[51] can function as an all-optical THz photonic synapse. The distinct positive and negative $\Delta T/T_0$ states can be employed for writing and erasing operations, respectively. This process can be controlled using a shutter and a rotator. Initially, with the shutter closed, no excitation occurs. When the shutter is opened, an optical pulse excites the sample at 0-degree sample azimuth, resulting in a positive $\Delta T/T_0$ signal, which is then converted into an electrical signal and retained by the connected memory device, effectively performing a writing operation [Figure 4(a)]. Since the memory device maintains the previous state, successive exposures to the optical pulse continue to build upon the stored signal, analogous to optical writing. To reduce or erase the accumulated signal, the shutter is used in a similar manner, but with the sample rotated 90° about its azimuthal axis. In this orientation, $\Delta T/T_0$ exhibits a negative response. When converted into an electrical signal, this negative output facilitates the reduction of the already built signal in the memory device, effectively enabling an erasure operation [Figure 4(b)]. Asymmetry in the strength of positive and negative THz signals will be overcome using op-amp comparator (zero-crossing detector). Overall, with continued technological progress, the proposed arrangement [Figure 4(c)], although presently at a conceptual stage, represents a promising route toward enabling both writing and erasing operations using the same set of optical and THz radiation. Such an approach can facilitate the development of more efficient and versatile THz-based artificial photonic synapses.

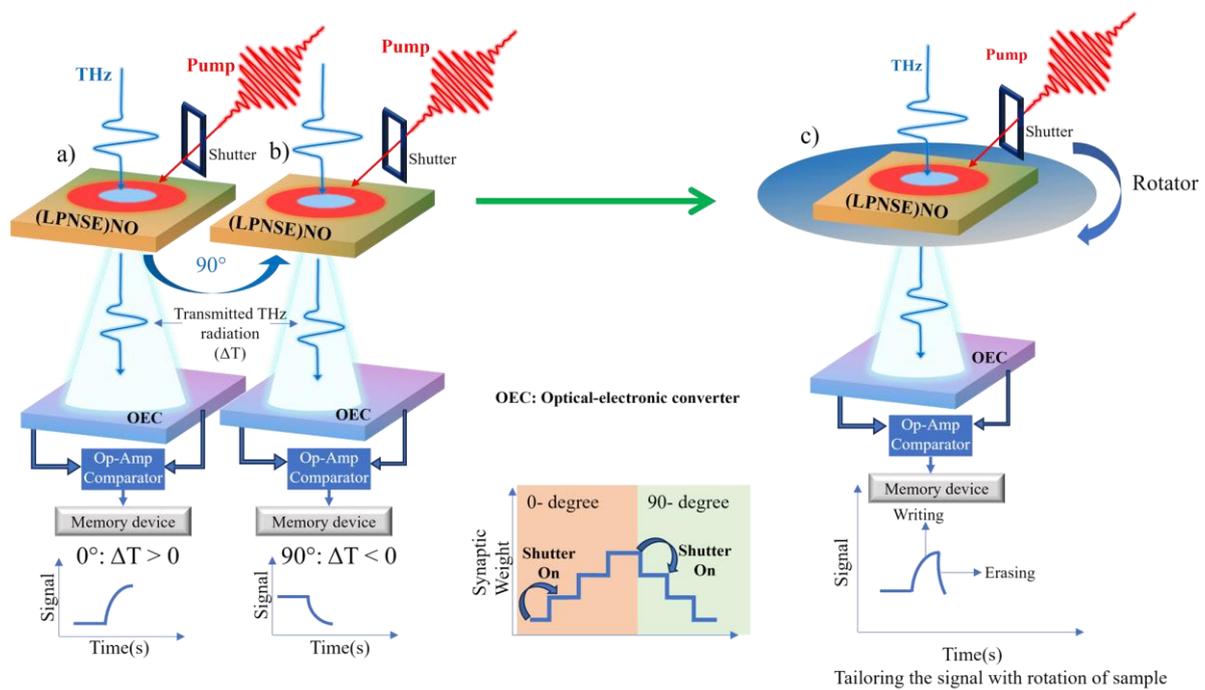

*Figure 4: Schematic depicting a proposed mechanism to develop a THz based artificial photonic synapse. Device with sample at (a) 0°(writing) and (b) 90° (erasing). (c) Illustration of working of the proposed device as a whole.*

## Summary

In conclusion, we demonstrated that combining emergent high entropy attribute with epitaxial engineering in rare earth nickelate can potentially generate novel ultrafast switching properties, which can potentially enhance efficiency of artificial photonic synapses for brain inspired computing. This study highlights 3 key facets, i) role of high entropy, ii) role of crystallographic epitaxial engineering and iii) high resolved ultrafast attributes of THz techniques. Unification of all these aspects on a single platform can be dealbreaker for new avenues in neuromorphic computing applications.

## Experimental and Methods Section

*Thin film synthesis:* Epitaxial (LPNSE)NO thin films were grown on single-crystal substrates using the pulsed laser deposition (PLD) from Excel Instruments employing a KrF excimer laser of wavelength 248 nm (Compex Pro from Coherent). Key growth parameters - such as the energy density of the KrF excimer laser falling on the polycrystalline target of (LPNSE)NO, substrate temperature, and oxygen partial pressure - were carefully optimized to achieve high-quality films.

*THz-TDS:* Tera K-8 from Menlosystems with an integrated cryostat was used to undertake temperature dependent as well as room temperature THz-time domain measurements.

*Optical Pump THz Probe (OPTP) spectroscopy*: Home-built spectrometer was utilized to measure transient change in THz transmission upon excitation by 800 nm femtosecond pulse. More details about the spectrometer are available in Supplementary Information, S9.

*DFT calculations:* First-principles calculations were performed using density functional theory (DFT) within the projector augmented wave formalism, as implemented in Vienna Ab initio Simulation Package (VASP),[52] for the room temperature orthorhombic phase of $La_{0.2}Pr_{0.2}Sm_{0.2}Nd_{0.2}Eu_{0.2}NiO_3$. Exchange-correlation functional was treated using the generalized gradient approximation parameterized by Perdew, Burke, and Ernzerhof.[53] The plane-wave cutoff and the energy convergence criterion were set to 450 eV and $10^{-7}$ eV, respectively. The Brillouin zone was sampled using a $\Gamma$-centred $5 \times 5 \times 3$ $k$-mesh. Electron correlation effects in Ni $3d$ orbitals were accounted for using the DFT+$U$ approach, with an effective on-site Coulomb interaction ($U$) value of 2 eV. The disordered rare-earth site was modelled using the virtual crystal approximation (VCA) [3], in which each rare-earth site is replaced by an effective atom having an averaged potential and occupancy. Oxygen vacancies were incorporated within the same VCA framework,[54] providing a computationally efficient approach to describe vacancy-induced effects.

## Supplementary Information

See the supplementary information for additional information that supports the findings of this work.

## Acknowledgement

D.S.R. thanks the Science and Engineering Research Board, Department of Science and Technology, New Delhi, for financial support under research Project No. CRG/2020/002338. S.K. thanks Council of Scientific and Industrial Research for financial support. B.S.M thanks Prime Minister Research Fellowship (PMRF; 0401968) funding agency, Ministry of Education, New Delhi. J.S. acknowledges the DST-INSPIRE for the fellowship (No. DST/INSPIRE/03/2018/000699).

## Conflict of Interest

Authors declare no conflict of interest.

## Data Availability Statement

The data that support the findings of this study are available from the corresponding author upon reasonable request.

# Room-Temperature Terahertz Photoconductivity Polarity Switching in High Entropy Nickelates with Implications for Photonic Synapses

Sanjeev Kumar, Brijesh Singh Mehra, Gaurav Dubey, Prakhar Vashishtha, Neeraj Bhatt, Jayaprakash Sahoo, Ravi Shankar Singh and Dhanvir Singh Rana[*]

S. Kumar, B.S. Mehra, G. Dubey, P. Vashishtha, N. Bhatt, J. Sahoo, R.S. Singh, D.S. Rana
Department of Physics
Indian Institute of Science Education and Research Bhopal, Bhopal 462066, Madhya Pradesh, India
E-mail: dsrana@iiserb.ac.in

## S1. Structural characterization of thin films

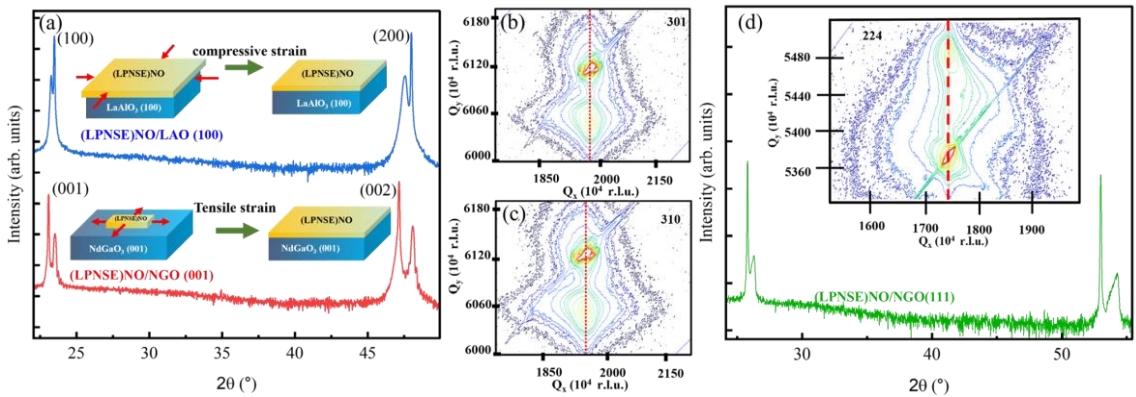

*Figure S1: Structural characterization of thin films. (a) XRD scans of (LPNSE)NO/NGO(001) and (LPNSE)NO/LAO(100). (b) and (c) Representative RSM scans of (LPNSE)NO/LAO(100) along (301) and (310) asymmetric directions respectively. (d) XRD scan of (LPNSE)NO/NGO(111), inset shows its RSM scan along (224) asymmetric direction.*

The θ-2θ XRD patterns of the (LPNSE)NO films grown on LaAlO$_3$ (100), NdGaO$_3$ (001) and NdGaO$_3$ (111)  substrates exhibit sharp and intense peaks corresponding only to the film and substrate, with no additional peaks or shoulders [Figure S1]. This confirms that the films are phase-pure, single-crystalline, and epitaxially oriented along the out-of-plane direction. The absence of secondary phases or broadened reflections rules out the presence of polycrystalline grains or misoriented domains.

Furthermore, the reciprocal space maps (RSMs) collected around asymmetric reflections [Figure S1 (b), (c) and inset of S1 (d)] provide direct evidence of single crystalline nature of film. In all the RSMs, the film and substrate diffraction peaks are sharply defined and well aligned along the in-plane reciprocal axis ($Q_x$), suggesting that the in-plane lattice parameters of the films are coherently matched to those of the substrates. This demonstrates that the films are fully strained, maintaining coherent epitaxy.

Importantly, the RSM data show single, symmetric diffraction spots without any peak splitting, broadening, or diffuse scattering features. If significant mosaicity, dislocations, or grain like morphology, one would expect to observe streak-like diffuse scattering or multiple film peaks corresponding to slightly misoriented crystalline domains. The absence of such features in our RSMs strongly supports that the films possess uniform crystallographic orientation and very low defect density.

**S2.**

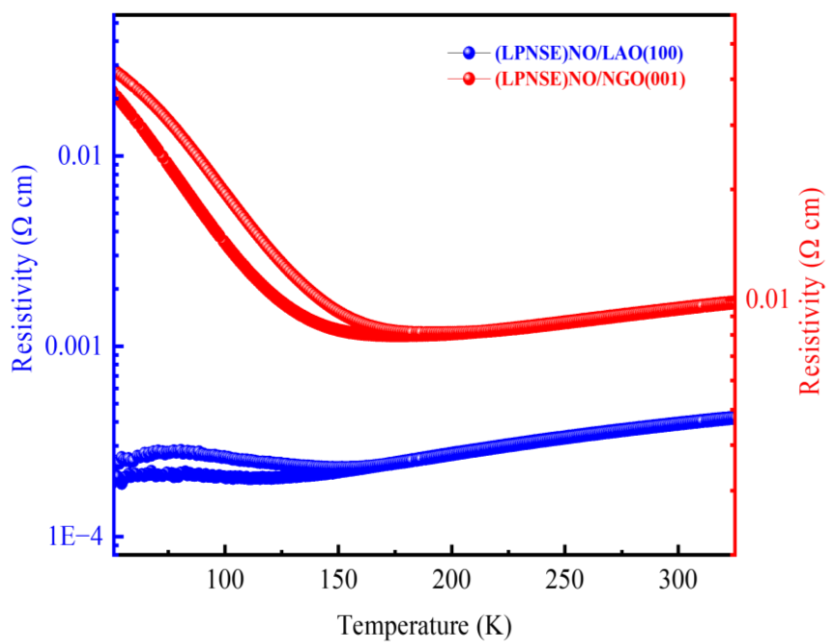

*Figure S2: Electronic characterization of thin films of (LPNSE)NO/LAO(100) and (LPNSE)NO/NGO(001).*

**S3.**

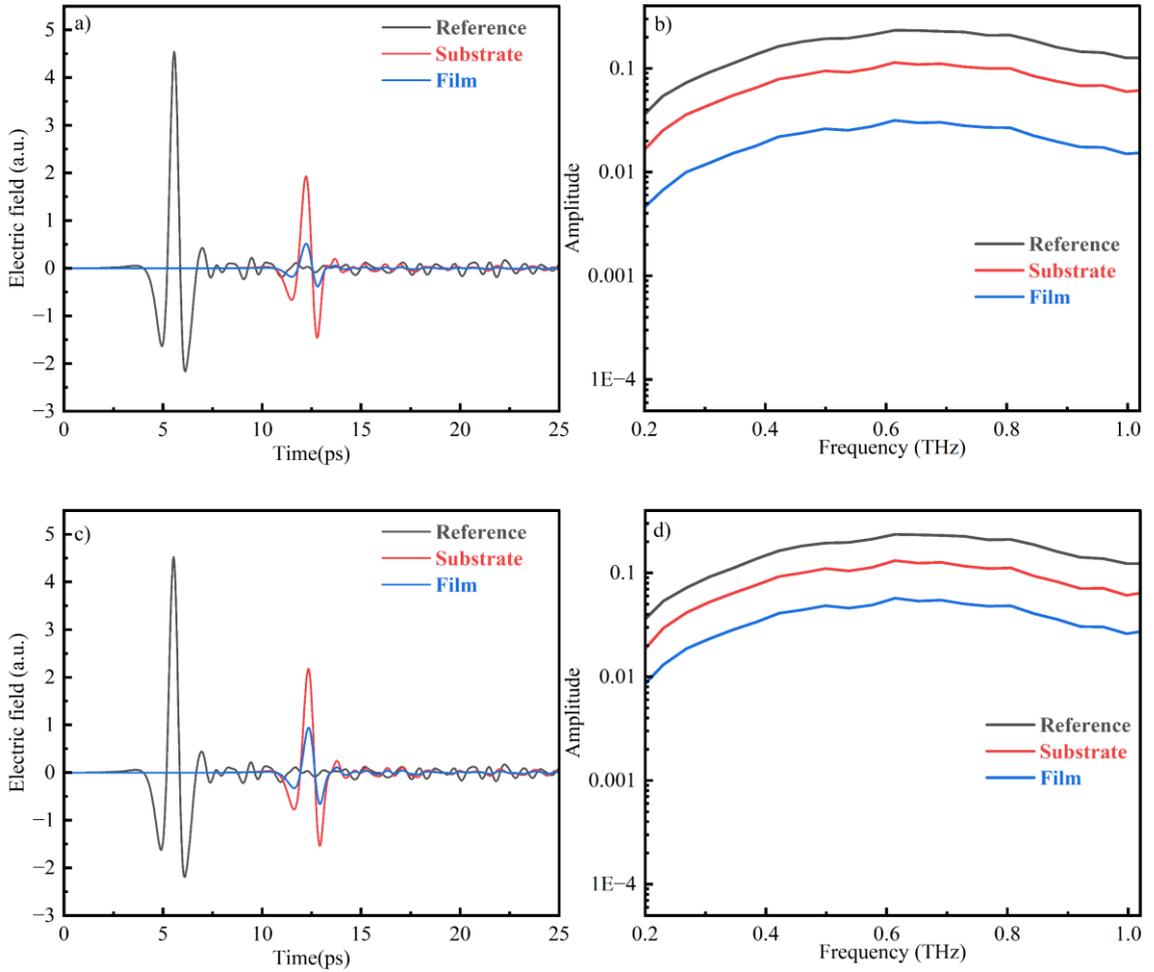

*Figure S3: Representative raw THz data. (a) THz time domain waveforms and (b) corresponding FFTs for (LPNSE)NO/LAO(100). (c) THz time domain waveforms and (d) corresponding FFTs for (LPNSE)NO/NGO(001).*

**S4.**

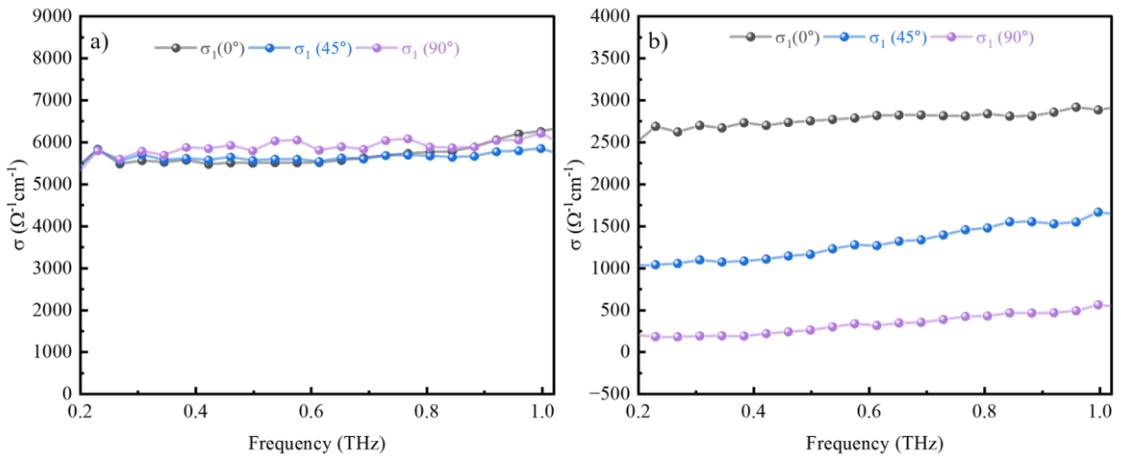

*Figure S4: Real THz conductivity at 0°, 45° and 90° for (a) (LPNSE)NO/LAO(100) and (b) (LPNSE)NO/NGO(001).*

**S5.**

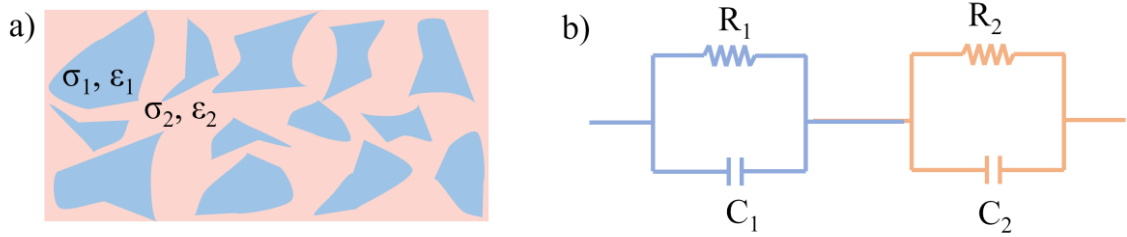

*Figure S5: Representation of a heterogenous system and (b) its equivalent circuit when the current flows through it.*

Current in these systems flows through regions with differing conductivities and dielectric constants, steady-state conditions lead to charge accumulation at the interface between these regions resulting in a Maxwell-Wagner dielectric response.

**S6.**

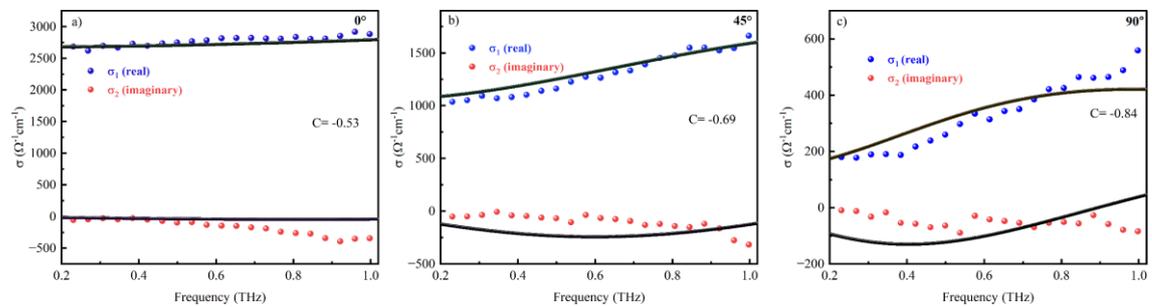

*Figure S6: Drude-smith fitting of THz conductivity along with obtained C parameter for (LPNSE)NO/NGO(001) at (a) 0°, (b) 45° and (c) 90°*

**S7.**

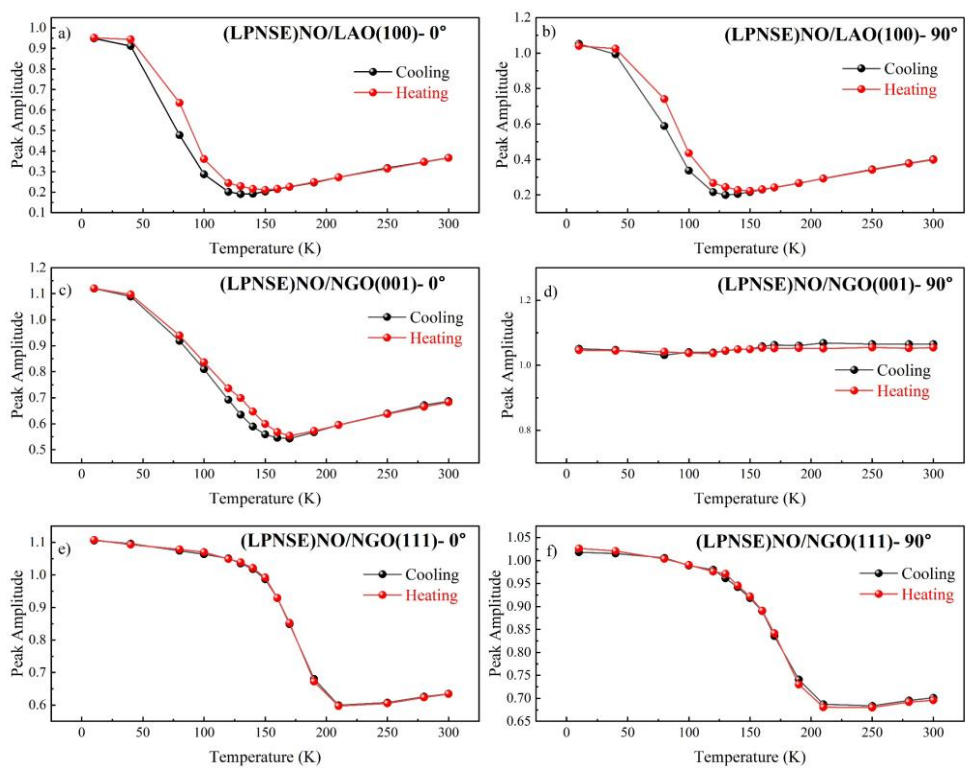

*Figure S7: Variation of peak amplitude as a function of temperature.*

**S8.**

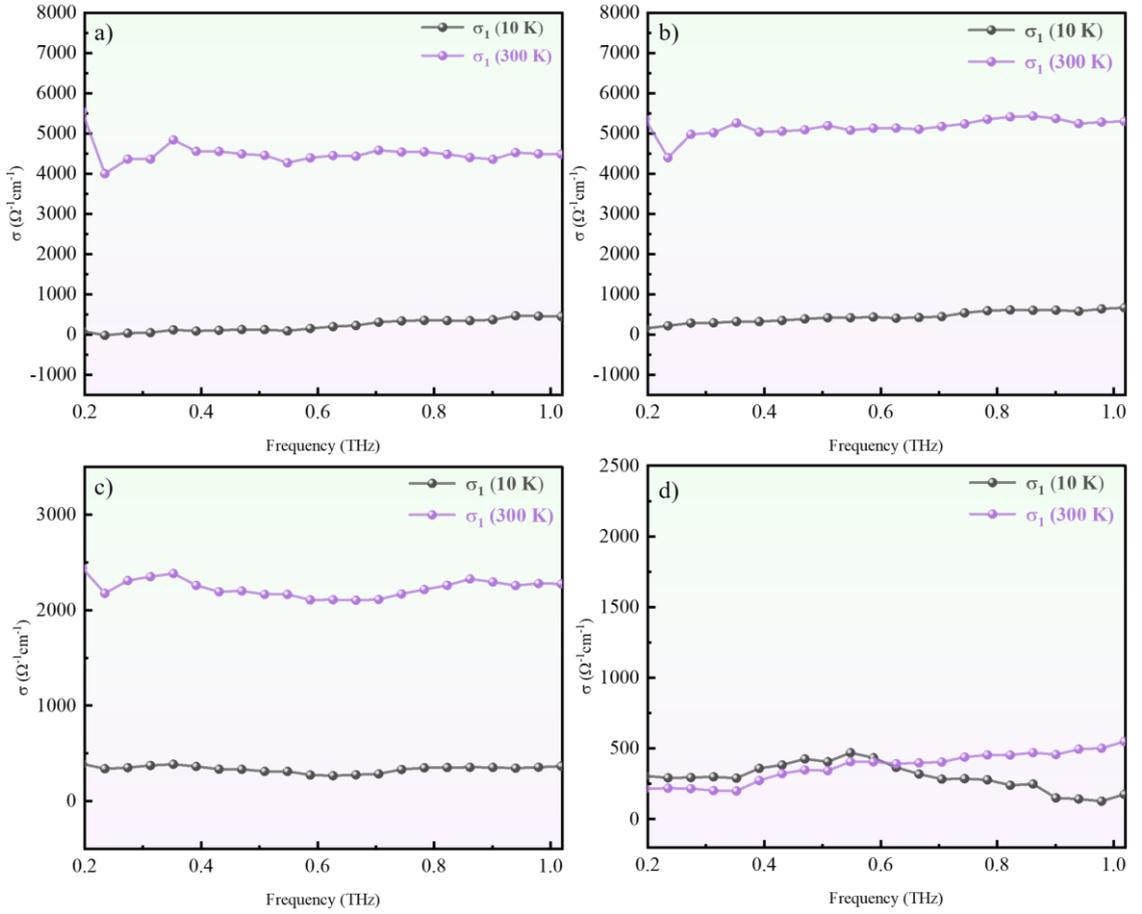

*Figure S8: Real THz conductivity for (LPNSE)NO/LAO(100) at 10 K and 300 K for (a) 0° and (b) 90° azimuthal angles. THz conductivity for (LPNSE)NO/NGO(001) at 10 K and 300 K for (c) 0° and (d) 90° azimuthal angles.*

## S9. Optical pump THz probe spectroscopy details

Laser beam (Pulse width: 35 fs, l: 800nm, Repetition Rate: 5KHz, Output power: 8W) was divided into two parts, one beam (80% of total intensity) was utilized to excite the sample, and another was further divided into two parts for generation and detection of THz radiation. {110} oriented ZnTe single crystals were used for generation of THz radiation by optical rectification and detection by electrooptic sampling. THz radiation was focused onto the sample using a pair of Off-Axis Parabolic mirrors and optical pump beam was incident obliquely onto the sample. Change in the THz transmission was measured as a function of time delay between pump and probe pulses, time delay between pump and probe pulses was varied using optical delay stages which convert spatial movement into temporal delay with a minimum possible time delay of ~6 fs.

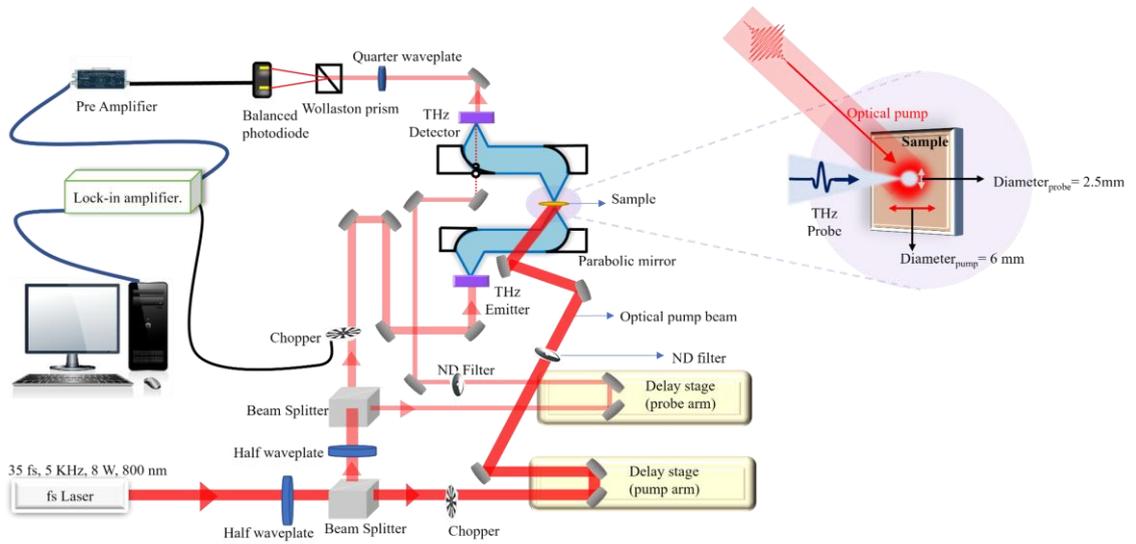

*Figure S9: Schematic of setup employed for carrying out optical pump THz probe spectroscopy. The optical pump beam illuminates a circular region with a diameter of approximately 6 mm, while the THz probe beam is focused onto the sample over a smaller circular area of less than 2.5 mm in diameter (inside 6 mm of pump illuminated part).*

**S10.**

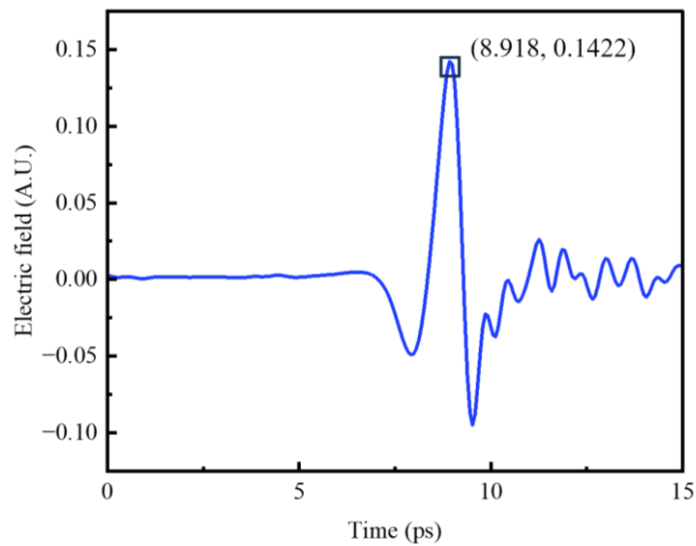

*Figure S10(a): Representative figure showing peak value of THz electric field (0.1422) and the corresponding time position (8.918 ps).*

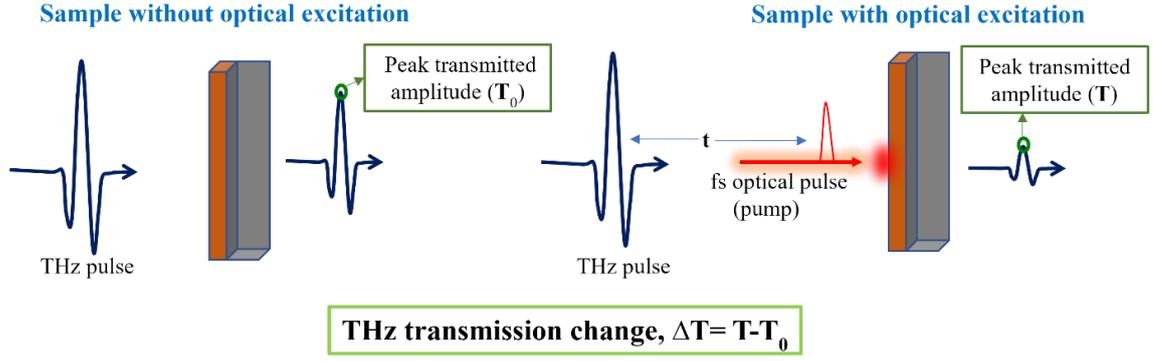

*Figure S10(b): Scheme of measurement for measuring ΔT.*

## S11.

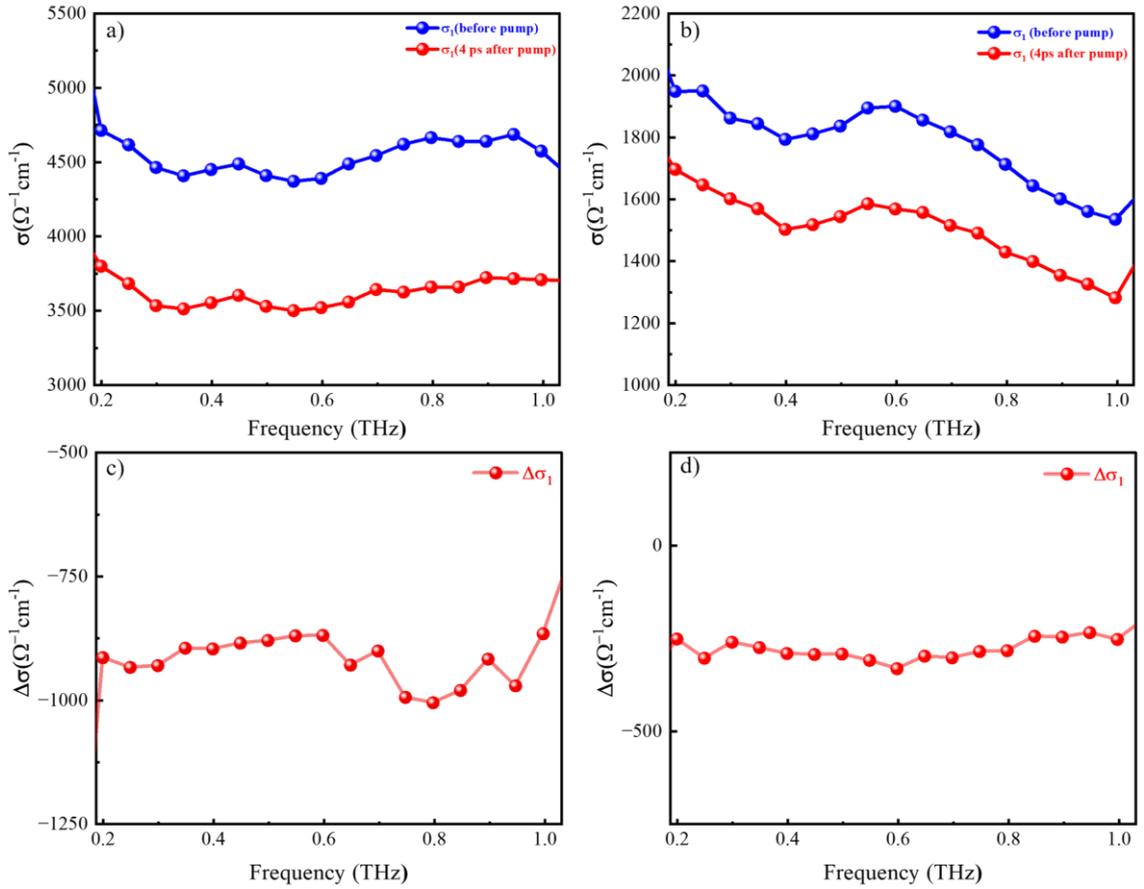

*Figure S11: Real THz conductivities obtained at different excitation conditions for (a) (LPNSE)NO/LAO(100) and (b)(LPNSE)NO/NGO(001). Change in real THz conductivity(photoconductivity) obtained 4 ps after the optical excitation for (c) (LPNSE)NO/LAO(100) and (d) (LPNSE)NO/NGO(001).*

**S12.**

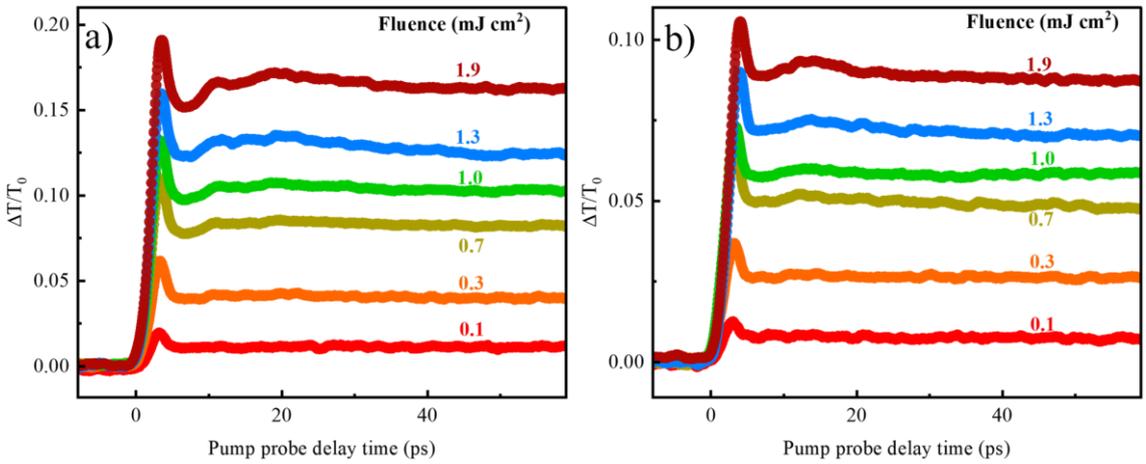

*Figure S12: Transient THz transmission change. (a) and (b) illustrates transient THz transmission as a function of pump probe delay time at different fluences for (LPNSE)NO/LAO(100) and (LPNSE)NO/NGO(001) respectively.*

**S13.**

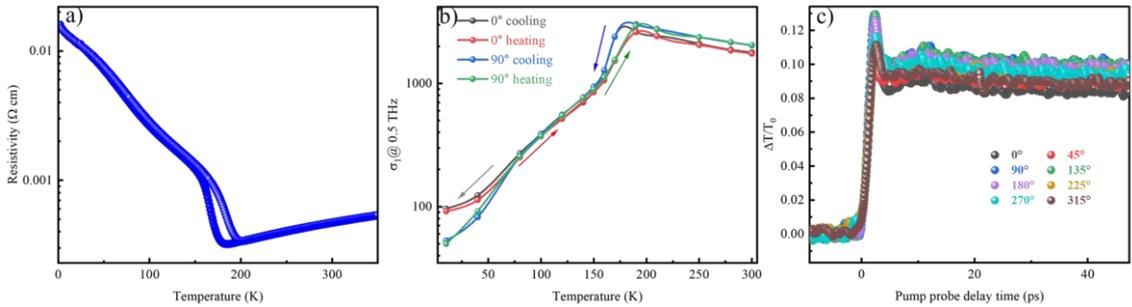

*Figure S13: Measurements on new low vacancy (LPNSE)NO/NGO(001) thin film. (a) Resistivity vs temperature plot. (b) Real conductivity at 0.5 THz at orthogonal sample azimuths as a function of temperature. (c) Normalized THz transmission for various sample azimuths.*

## S14. Quantitative experimental evidence for anisotropic ordering of oxygen vacancies

*(i) Variation of the 'C' parameter with azimuthal angle:* The systematic modulation of the 'C' parameter as a function of azimuthal rotation [Figure S14(i)] clearly points to a directional distribution of oxygen vacancies along specific crystallographic orientations. Since the 'C' parameter is known to be sensitive to factors such as grain boundaries, crystal defects, and oxygen vacancies, and considering that in the present case the films are of high crystalline quality—epitaxial and fully coherent—the influence of grain boundaries and other structural defects can be reasonably excluded.

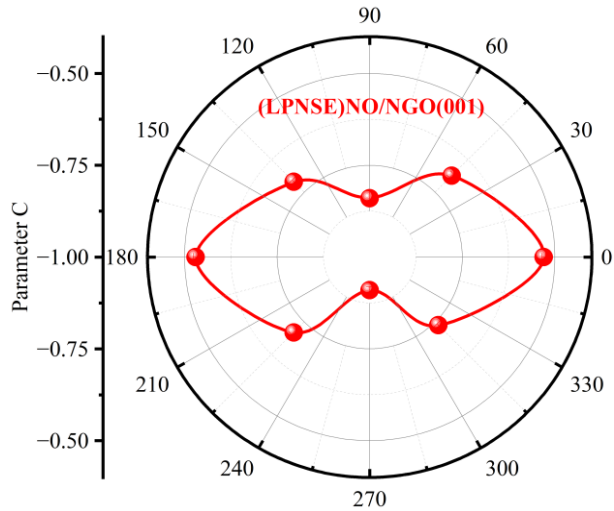

*Figure S14(i): Variation of C parameter with sample azimuth for (LPNSE)NO/NGO(001)*

**(ii) Anisotropy in the metal–insulator transition behavior in THz conductivity:** The observed crossover—from a clear MIT behavior to its suppression upon 90° rotation—indicates that oxygen vacancies possess a preferential alignment rather than a random distribution, leading to the emergence of directional transport anisotropy. Specifically, at room temperature, the conductivity along the direction orthogonal to the primary MIT axis is lower, consistent with enhanced carrier scattering from oxygen vacancies acting as localized defect centers. However, upon cooling, the conductivity along this axis becomes relatively higher—a trend that can be understood as oxygen vacancies acting as donor and decreasing the film's resistivity.[1] Confirming 90° orientation has higher concentration of oxygen vacancies.

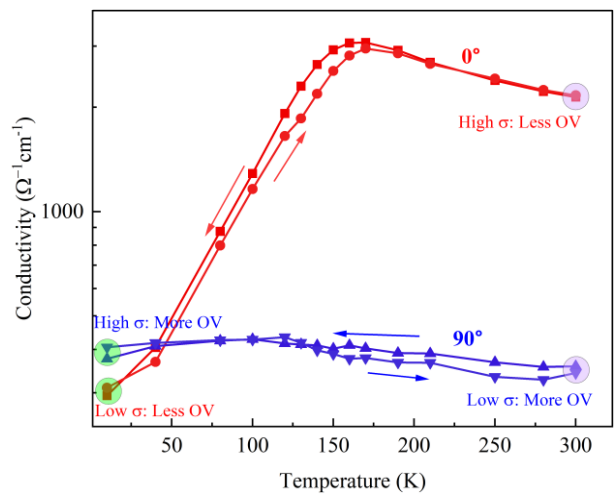

*Figure S14(ii): Temperature dependent real THz conductivity at 0.5 THz along orthogonal crystal axes for (LPNSE)NO/NGO(001).*

**(iii) Disappearance of anisotropy with varying oxygen vacancy concentration:** By systematically tuning the oxygen vacancy concentration—verified through room-temperature resistivity measurements—we observed a gradual reduction and eventual disappearance of the anisotropic features in the THz response. To show this experimentally, we synthesized a new (LPNSE)NO/NGO(001) film with a lower concentration of oxygen vacancies, where the $\Delta T/T_0$ signals at different azimuthal angles does not exhibit any significant anisotropy as opposed to old (LPNSE)NO/NGO(001), which had higher concentration of oxygen vacancies [Figure S14(iii)(c)]. This strong correlation between oxygen stoichiometry and anisotropy provides direct experimental evidence of the pivotal role played by oxygen vacancies toward anisotropic THz photoconductivity.

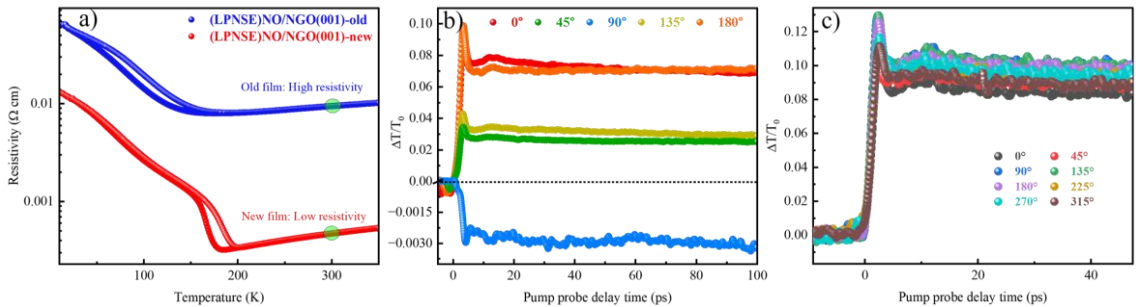

*Figure S14(iii): (a) Temperature dependent resistivity for old and new (LPNSE)NO/NGO(001) films, new film has lesser concentration of oxygen vacancies as also evident from lower resistivity at room temperature. (b) and (c) illustrates transient THz transmission at various sample azimuths for old and new films respectively*

## S15.

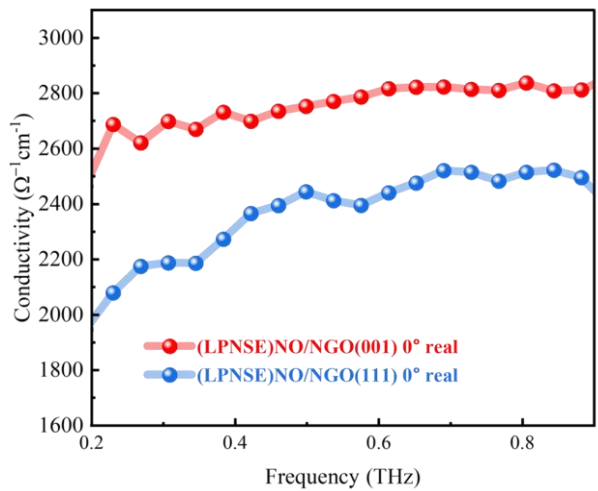

*Figure S15: Real THz conductivity of (LPNSE)NO thin film on NGO(001) and NGO(111) at 0° azimuthal angle.*

## S16. Comparison high entropy nickelate with parent rare earth nickelate (NdNiO$_3$)

The compositional complexity in high entropy materials introduce significant configurational disorder and entropic stabilization, leading to novel structural and electronic characteristics that are absent in conventional single-element perovskite nickelates. Moreover, high entropy nickelates exhibit enhanced mechanical stability compared to their parent compounds, making them promising candidates for the development of robust and durable device architectures. In HEN, the A-site disorder can promote competing interactions, which can enhance and readily enable both negative and positive THz photoconductivity responses. Importantly, this study constitutes the *first ever systematic investigation of high entropy nickelates in the THz regime*, thereby establishing a new framework for probing their ultrafast optical and electronic dynamics. Table below elucidates the differences based on some key features between parent rare earth nickelate (NdNiO$_3$) and high entropy nickelate ((La$_{0.2}$Pr$_{0.2}$Nd$_{0.2}$Sm$_{0.2}$Eu$_{0.2}$) NiO$_3$).

### Table S1

| Feature | NdNiO$_3$ | (La$_{0.2}$Pr$_{0.2}$Nd$_{0.2}$Sm$_{0.2}$Eu$_{0.2}$) NiO$_3$ (High Entropy) |
|---|---|---|
| Structural disorder | Minimal | High entropy/stochastic mixing |
| MIT Control/Resilience | Sensitive, strain/doping-dependent | Robust, averaged and strain-tunable |
| Orbital polarization | Typical for single-ion compound | Enhanced, resilient with disorder |
| Application flexibility | Moderate, conventional | High, tunable for advanced functions |

*Advantages of this material system*

A key advantage of this material platform lies in its inherently high compositional flexibility, arising from the presence of five distinct elements at the A-site enabling tunability of properties. By judiciously modifying the A-site dopant element and their ratios, one can potentially tailor, enhance, or even induce entirely new THz functionalities that are unattainable in conventional single-cation nickelates. As a result, high entropy nickelates offer a versatile pathway for expanding the functional landscape of the nickelate family. In the present work, we have focused primarily on the THz response under optical excitation and varying temperatures. However, it is reasonable to anticipate that additional external perturbations—such as strain, electric fields, or magnetic fields—can reveal further emergent and technologically relevant behaviours. Moreover, within the context of the present study, the magnitude of both positive and negative THz signals can be deliberately engineered, thereby improving the operational efficiency of the proposed artificial photonic synapse. These prospects underscore the vast potential of high entropy nickelates for the development of multifunctional materials and next-generation adaptive devices, which could ultimately play a pivotal role in future artificial intelligence–driven technologies.